\documentclass[aps,prl,reprint,superscriptaddress]{revtex4-1}

\usepackage[pdftex]{graphicx}
\usepackage{sidecap}
\usepackage{color}
\usepackage{ragged2e}

\begin{document}

\graphicspath{{.}{figures/}}

\title{Strong $(\pi,0)$ spin fluctuations in $\beta$-FeSe observed by neutron spectroscopy}

\author{M. C. Rahn}
\email[]{marein.rahn@physics.ox.ac.uk}
\affiliation{Department of Physics, University of Oxford, Clarendon Laboratory, Oxford, OX1 3PU, United Kingdom}

\author{R. A. Ewings}
\affiliation{ISIS Facility, STFC Rutherford Appleton Laboratory, Harwell Oxford, Didcot, OX11 0QX, United Kingdom}

\author{S. J. Sedlmaier}
\author{S. J. Clarke}
\affiliation{Department of Chemistry, University of Oxford, Inorganic Chemistry Laboratory, Oxford, OX1 3QR, United Kingdom}

\author{A. T. Boothroyd}
\email[]{a.boothroyd@physics.ox.ac.uk}
\affiliation{Department of Physics, University of Oxford, Clarendon Laboratory, Oxford, OX1 3PU, United Kingdom}



\date{\today}

\begin{abstract}
We have performed powder inelastic neutron scattering measurements on the unconventional superconductor $\beta$-FeSe ($T_{\rm c} \simeq 8$\,K). The spectra reveal highly dispersive paramagnetic fluctuations emerging from the square-lattice wave vector $(\pi,0)$ extending beyond 80\,meV in energy. Measurements as a function of temperature at an energy of $\sim$13\,meV did not show any variation from $T_{\rm c}$ to 104\,K. The results show that FeSe is close to an instability towards $(\pi,0)$ antiferromagnetism characteristic of the parent phases of the high-$T_{\rm c}$ iron arsenide superconductors, and that the iron paramagnetic moment is neither affected by the orthorhombic-to-tetragonal structural transition at $T_{\rm s} \simeq 90$\,K nor does it undergo a change in spin state over the temperature range studied.
\end{abstract}

\pacs{74.25.Ha, 74.70.Xa, 75.40.Gb, 78.70.Nx}

\maketitle

Iron selenide ($\beta$-Fe$_{1+x}$Se, hereafter denoted ``FeSe'') is structurally the simplest of the iron-based superconductors but it is also one of the most intriguing. The superconducting transition temperature of the pure bulk phase is relatively low, $T_{\rm c}\approx 8$\,K \cite{HSU08}, but it increases to 37\,K under pressure \cite{MED09} and rises above 40\,K with intercalation of alkali ions $A^+$ to form $A_x$Fe$_{2-y}$Se$_2$ \cite{GUO10} or by co-intercalation of ammonia molecules and amide ions or organic molecules along with $A^+$ \cite{YIN12,BUR13,KRZ12}. Very recently, superconductivity was reported at temperatures as high as 100\,K in monolayers of FeSe on SrTiO$_3$ \cite{WAN12,GE14}.  Although there is evidence that superconductivity at ambient pressure is favored by reduction of Fe below the +2 oxidation state and minimisation of vacancies in the FeSe layers \cite{BUR13,SUN15}, there is currently no simple explanation for such an extraordinary variation in $T_{\rm c}$ among derivatives containing very similar antifluorite layers of FeSe.

The structural and electronic ordering properties of FeSe differ qualitatively from those of the related iron pnictide compounds in two important ways. First, superconductivity appears in FeSe without the need for doping and is very sensitive to composition \cite{MCQ09_01}. Second, FeSe has a tetragonal-to-orthorhombic structural transition ($T_{\rm s} \simeq 90$\,K \cite{HSU08,MCQ09_02}), as in the parent phases of the iron pnictide superconductors, but this transition is not followed by the development of long-range magnetic order \cite{MIZ10}. The phase below $T_{\rm s}$ is considered to be some form of electronic nematic, but opinions divide over whether the nematic transition is driven by orbital ordering \cite{BAE14,NAK14,SHI14,BOE15} or by spin degrees of freedom \cite{CAO14,WAN15,GLA15,YU15}.

This paper reports measurements of collective paramagnetic spin fluctuations in FeSe. Spin fluctuations are a prominent feature of the iron-based superconductors and are thought to play a significant role in the pairing interaction \cite{HIR11,CHU12,SCA12}. In the iron arsenide superconductors, spin fluctuations emerge from the same (or nearly so) characteristic in-plane wave vector ${\bf q}_{\rm m} = (\pi,0)$, referred to the Fe square sub-lattice, as the spin density wave (SDW) order of the parent phases.  This magnetic instability is understood to be assisted by nesting of hole and electron Fermi surface pockets centred around the $\Gamma$ and X points of the square lattice. Spin fluctuations have also been observed in the superconducting iron selenides, but the characteristic wave vector varies from system to system. For example, it is $(\pi,0)$ in FeTe$_{1-x}$Se$_x$ ($x\approx 0.5$) \cite{QIU09}, $(\pi,\pi/2)$ in $A_x$Fe$_{2-y}$Se$_2$ ($A$ = K, Rb, Cs) \cite{FRI12,PAR11,TAY12}, and different again in Li$_{x}$(ND$_{2}$)$_{y}$(ND$_{3}$)$_{1-y}$Fe$_{2}$Se$_{2}$ \cite{TAY13}.

{\it Ab initio} electronic structure calculations indicate that FeSe is close to a magnetic ordering instability with characteristic wave vector $(\pi,0)$ \cite{SUB08,ESS12,HEI14}. However, angle-resolved photoemission spectroscopy and quantum oscillation studies have revealed that the Fermi surface deviates significantly from the predictions \cite{TER14,MAL14,NAK14,SHI14,AUD15,WAT15}, and several models for the nematic phase predict competing magnetic phases with ${\bf q}_{\rm m} = (\pi,\xi)$, $0 \le \xi \le \pi/2$ \cite{CAO14,WAN15,GLA15,YU15}. Experimental information on the magnetic ground state of FeSe is currently lacking, and is urgently needed to elucidate the nematic phase and to assess the role of spin fluctuations in the superconducting state.

Here we report observations of the wave vector and energy dependence of the spin fluctuations in FeSe by powder inelastic neutron scattering. We find collective spin fluctuations emerging from $(\pi,0)$ and equivalent square-lattice wave vectors, extending to energies greater than 80\,meV. We do not observe any significant change in the low energy ($\sim$10--15\,meV) part of the spectrum on crossing the orthorhombic-to-tetragonal transition.
 \begin{figure*}
\includegraphics[width=2\columnwidth,trim= 0pt 0pt 0pt 0pt, clip]{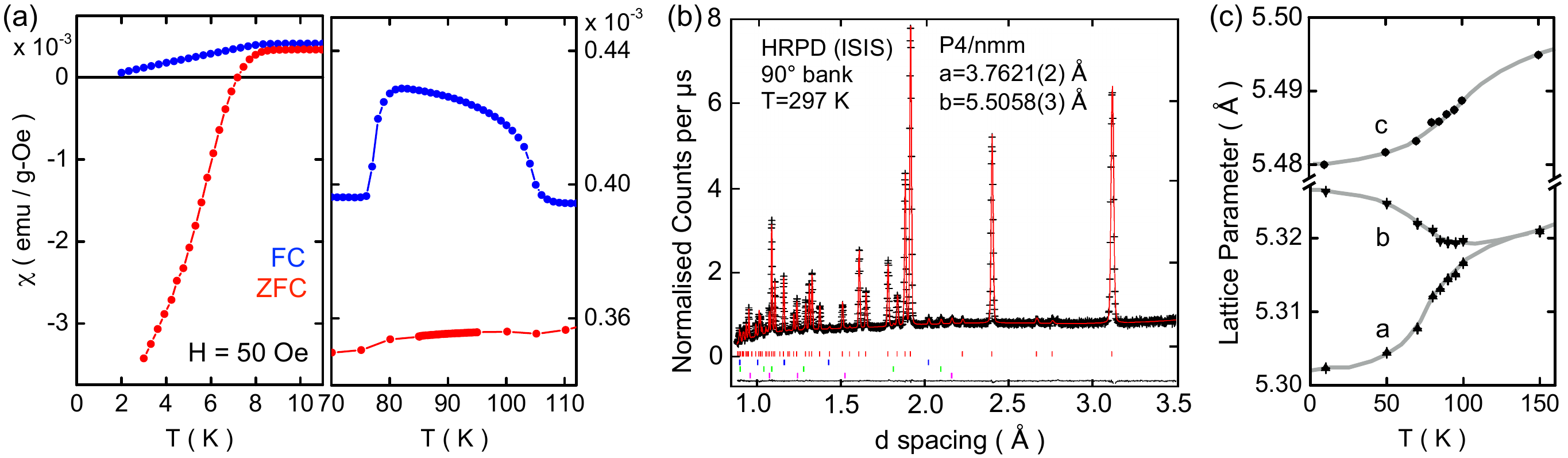}%
 \caption{\label{fig1}(color online). (a) Magnetic susceptibility of FeSe powder. The field-cooled (FC) and zero-field cooled (ZFC) curves confirm the onset of superconductivity at $T_{\rm c} \simeq 8$\,K (left). The tetragonal-to-orthorhombic structural transition at $T_{\rm s} \simeq 90$\,K is signalled by a broad magnetic anomaly (right). (b) Rietveld refinement against room temperature neutron powder diffraction data of FeSe. Peak positions for the $\beta$-FeSe phase are marked by vertical red ticks beneath the data. The other ticks indicate peak positions for Fe impurities and the vanadium sample can. (c) Temperature dependence of the orthorhombic lattice parameters of FeSe. The points at 150\,K are the tetragonal parameters with $a$ multiplied by $\sqrt{2}$. The lines are visual guides.}
 \end{figure*}


A powder sample of FeSe of total mass 13.8\,g was prepared in five separate batches of 2--3\,g each. All handling was carried out in an argon atmosphere. Iron and selenium powders (5N purity) were ground together, sealed under vacuum in a silica glass ampoule and reacted at 700$^{\circ}$C for 24\,h. The product of this reaction was reground, resealed under vacuum, annealed at 700$^{\circ}$C for 38\,h and then cooled to 400$^{\circ}$C and held for 6 days.  The ampoule was then quenched in ice water and the sample ground to a fine powder. The batches were found to be of very high phase purity by x-ray and neutron diffraction, with trace amounts ($<1$\%) of hexagonal $\alpha$-FeSe and unreacted Fe as the only detectable impurities.

Magnetisation measurements, performed with a Superconducting Quantum Interference Device (SQUID) magnetometer, confirmed the onset of superconductivity at $T_c \simeq 8\,\mathrm{K}$ in each of the five batches. An example of field-cooled and zero-field-cooled data is shown in Fig.~\ref{fig1}(a). Measurements retaken after the neutron scattering experiment confirmed that the sample did not deteriorate.  The right-hand panel of Fig.~\ref{fig1}(a) shows a broad magnetic anomaly at the structural transition $T_{\rm s}\simeq 90\,K$ consistent with previous data on FeSe powders \cite{HSU08}.

 For a detailed structural analysis, we performed high resolution neutron powder diffraction on the HRPD instrument at the ISIS Facility.  Measurements were made at temperatures between 10\,K and room temperature. Figure~\ref{fig1}(b) shows data collected at room temperature together with a Rietveld fit. The temperature dependence of the lattice parameters obtained from the refinements are shown in Fig.~\ref{fig1}(c).  The continuous tetragonal ($P4/nmm$) to orthorhombic ($Cmma$) transition at $T_s\simeq 90$\,K is consistent with earlier results \cite{HSU08,MAR08,MCQ09_02}. The orthorhombic distortion $(b-a)/a$ approaches $0.5\,\%$ at $10\,\mathrm{K}$. Refinement of the composition $\mathrm{Fe}_{1+x}$Se against data above and below $T_s$ yielded $x = 0.01(1)$, i.e.~with interstitial Fe sites between the stoichiometric FeSe layers occupied at the 1\% level, a finding consistent with a previous report correlating composition with $T_{\rm c}$ \cite{MCQ09_01}.

 Inelastic neutron scattering was performed on the chopper spectrometer MERLIN at the ISIS Facility \cite{MER06}. The powder sample was loaded into aluminium foil packets and placed in an aluminium can in annular geometry. The can was attached to a closed-cycle refrigerator. Neutron spectra were recorded with incident energies of $E_{\rm i} =34$, 50 and 100\,meV at temperatures from 8 to 104\,K.  The spectra were normalised to the incoherent scattering from a standard vanadium sample measured with the same incident energies, enabling us to present the data in absolute units of $\mathrm{mb}\,\mathrm{sr}^{-1}\,\mathrm{meV}^{-1}\mathrm{f.u.}^{-1}$ (where f.u. refers to one formula unit of FeSe).


 \begin{figure}
 \includegraphics[width=0.8\columnwidth,trim= 0pt 0pt 0pt 0pt, clip]{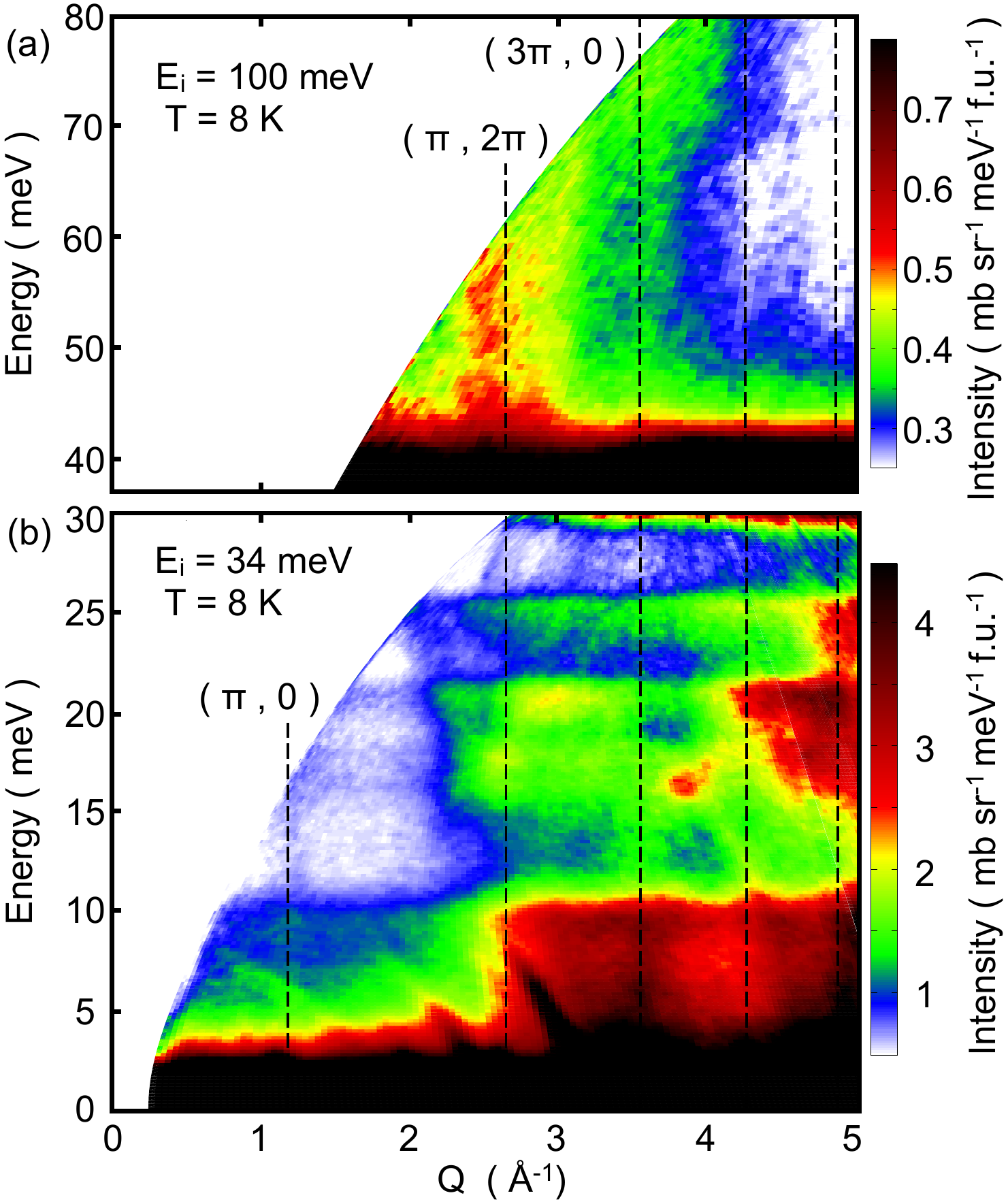}%
 \caption{\label{fig2} (color online). Neutron powder spectra of FeSe obtained on MERLIN at $T=8$\,K $\simeq T_{\rm c}$. The vertical dashed lines show the 2D wave vector $(\pi,0)$ and equivalent positions. (a) High-energy part of the spectrum recorded with an incident energy $E_{\rm i}=100$\,meV. The vertical bands of scattering above the phonon cut-off at 40\,meV are caused by steeply dispersing cooperative paramagnetic fluctuations. (b) Low energy part of the spectrum from the data measured with $E_{\rm i}=34$\,meV. Magnetic scattering is visible in the energy window 10--15\,meV where phonon scattering is weak.}

 \end{figure}

  \begin{figure}
 \includegraphics[width=0.8\columnwidth,trim= 0pt 0pt 0pt 0pt, clip]{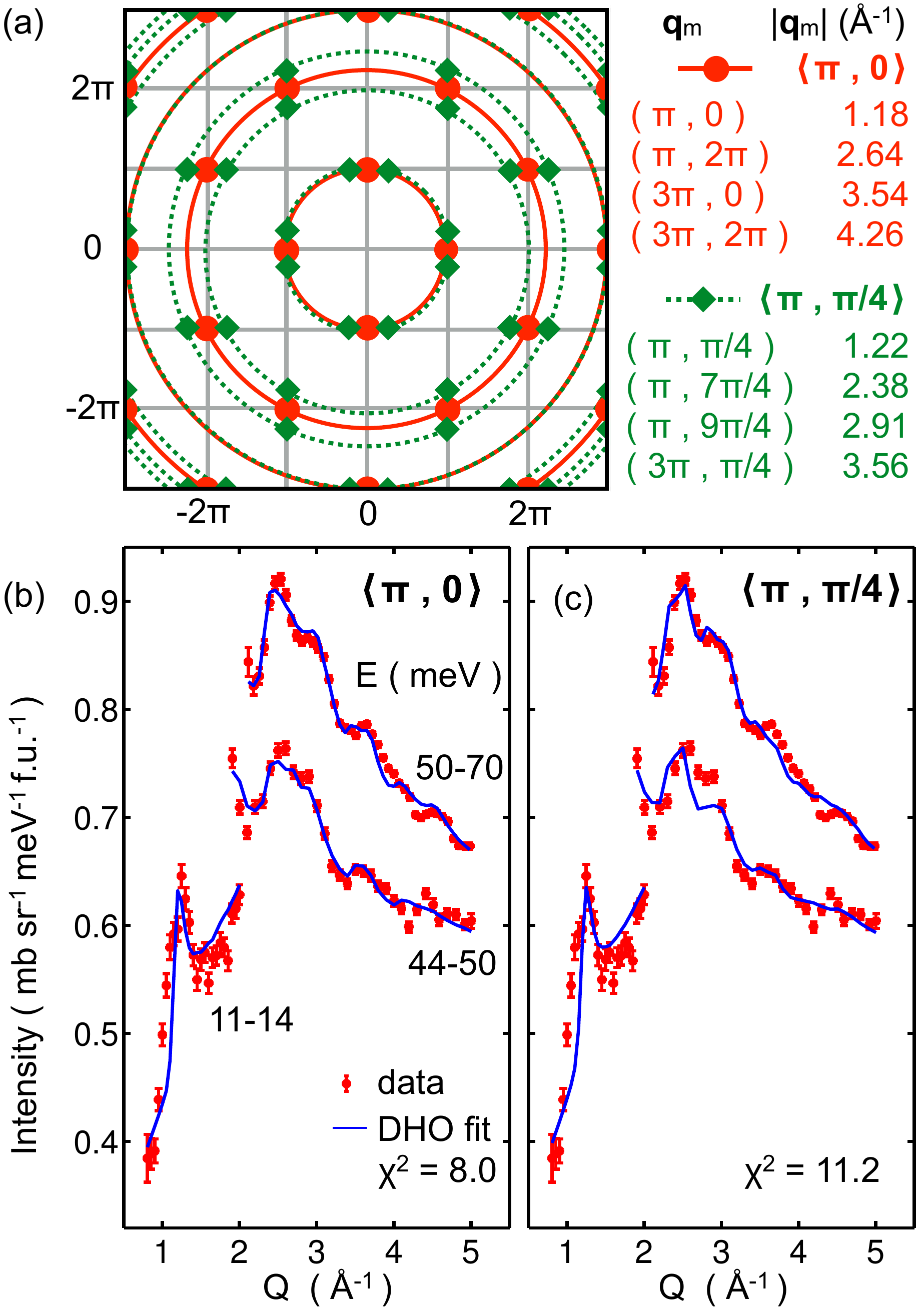}%
 \caption{\label{fig3} (color online). (a) Reciprocal lattice of the Fe square lattice. The (red) filled circles mark the $(\pi,0)$-type wave vectors and the (green) filled diamonds the $(\pi,\pi/4)$-type positions, including the equivalent 90$^{\circ}$ domains. The full and dotted circles show the effect of powder averaging, and the table lists the corresponding values of $Q=|{\bf q}_{\rm m}|$. (b) Constant-energy cuts through the data in Fig.~\ref{fig2} for three different energy bands as indicated. The upper two cuts are offset vertically. The (red) symbols are the data and the (blue) lines are fits with a damped harmonic oscillator model for spin-waves dispersing anisotropically from ${\bf q}_{\rm m}= (\pi,0)$. (c) The same as in (b) but  with ${\bf q}_{\rm m} = (\pi,\pi/4)$ and an isotropic dispersion.}
 \end{figure}

 \begin{figure}
 \includegraphics[width=0.8\columnwidth,trim= 0pt 0pt 0pt 0pt, clip]{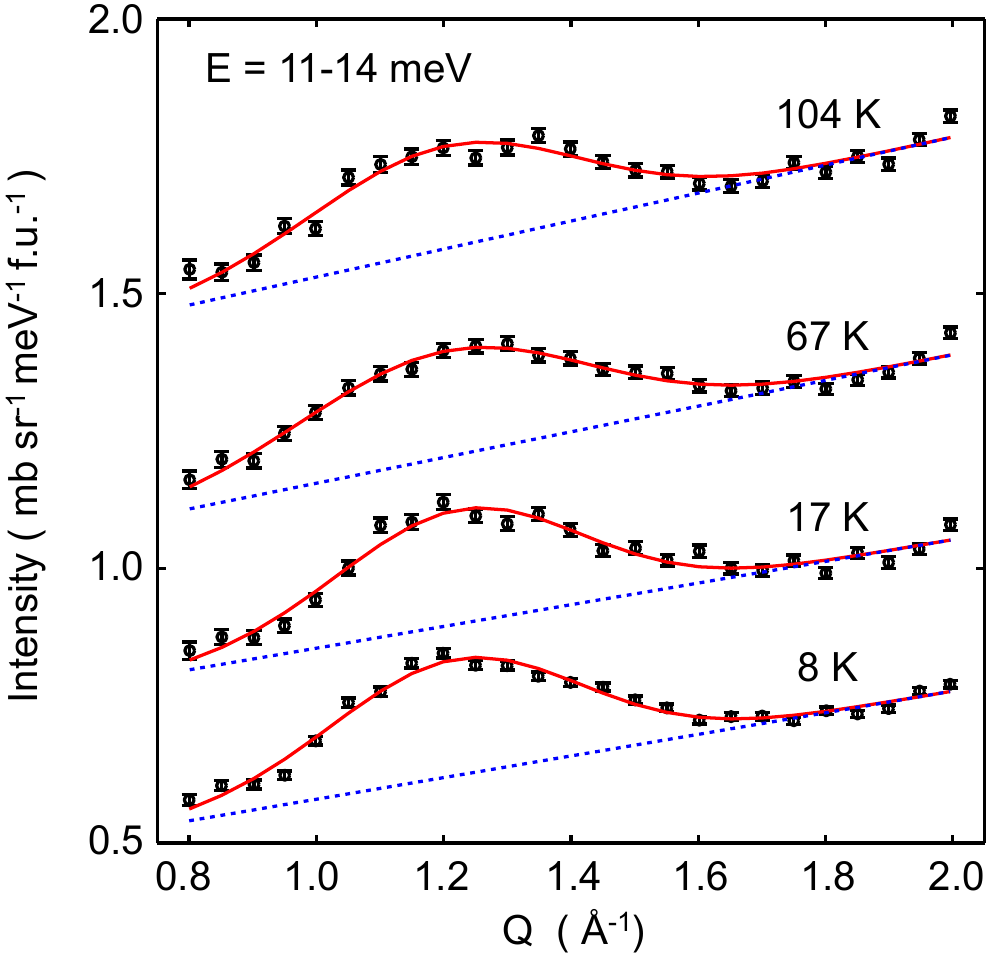}%
 \caption{\label{fig4} (color online). Temperature dependence of the magnetic scattering at $(\pi,0)\approx 1.2$\,{\AA}$^{-1}$ averaged over the energy range $11 - 14$\,meV. The solid lines are fits to a Gaussian function on a linear background (dotted). The upper three scans are offset vertically by 0.25, 0.5 and 0.75 units, respectively.}
 \end{figure}

Figure~\ref{fig2}(a) shows an intensity map of part of the $E_{\rm i} =100\,\mathrm{meV}$ spectrum measured at 8\,K on MERLIN. The spectrum is dominated by scattering from phonons for energies $E$ below the phonon cut-off at 40\,meV \cite{PHE09}. Above 40\,meV, there is a broad vertical column of scattering centred on the wave vector $Q=2.6$\,{\AA}$^{-1}$, and a weaker column centred on 3.5\,{\AA}$^{-1}$. Figure~\ref{fig2}(b) is a similar intensity map measured with $E_{\rm i} = 34\,\mathrm{meV}$ to probe the low $(Q,E)$ part of the spectrum. Phonon scattering dominates in this regime, but there is a window between 10 and 15\,meV in which the phonon signal is small, and a vertical column of weak scattering can be seen centred near $Q = 1.2$\,{\AA}$^{-1}$. Such scattering columns are observed in neutron powder spectra of other iron-based superconductors and have been confirmed to arise from strongly dispersive spin fluctuations \cite{EWI08,CHR08,ISH09,TAY11}.

The magnetic signals identified in the intensity maps can be seen in more detail in the $Q$ cuts made at fixed average energy shown in Fig.~\ref{fig3}(b). The cuts contain peaks centered on $Q=1.2$, 2.6 and 3.5\,{\AA}$^{-1}$, and there are additional weak signals near $Q = 4.5$\,{\AA}$^{-1}$. The series of magnetic peaks can be indexed as orders of the square lattice wave vector $(\pi, 0)$, see Fig.~\ref{fig3}(a). In reality, the magnetic signal will extend in the out-of-plane direction, either as a diffuse rod of scattering if the correlations are quasi-two-dimensional or as a series of peaks if there are strong inter-layer correlations. Simulations of such types of out-of-plane scattering show that after powder averaging the peaks have a tail on the high $Q$ side but the maxima shift by only a small amount ($<0.06$\,{\AA}$^{-1}$) from the ideal two-dimensional wave vectors.

Although FeSe does not order magnetically, our results show that it has a strong magnetic response at $(\pi, 0)$ and equivalent positions which characterise the in-plane SDW order found in the parent phases of the iron arsenide superconductors. To quantify the analysis, we compare the data to a phenomenological model for the low energy response of a two-dimensional (2D) antiferromagnetically-correlated paramagnet. The model has been used previously to describe the low energy part of the spectrum of superconducting Ba(Fe$_{1-x}$Co$_x$)$_2$As$_2$ \cite{LES10}. The neutron scattering cross section may be written
\begin{equation}
\frac{{\rm d}^2\sigma}{{\rm d}\Omega{\rm d}E_{\rm f}} = \frac{k_{\rm f}}{k_{\rm i}}S({\bf Q},E),\label{eq:1}
\end{equation}
where $S({\bf Q},E)$, the magnetic response function, is the quantity presented here. For an isotropic paramagnet,
\begin{equation}
S({\bf Q},E) = \left(\frac{\gamma r_0}{2\mu_{\rm
B}}\right)^{\hspace{-2pt}2}\frac{1}{1-\exp(-\beta E)}\frac{2}{\pi}f^2({\bf Q})\chi''({\bf q},E),\label{eq:2}
\end{equation}
where $(\gamma r_0/2)^2 = 72.7$\,mb, $\beta = 1/k_{\rm B}T$, $f({\bf Q})$ is the magnetic form factor, and $\chi''({\bf q},E)$ is the absorptive part of the generalized susceptibility.  The low-energy magnetic excitations are envisaged as damped spin waves with a linear dispersion, and we use a harmonic oscillator model
\begin{equation}
\chi''({\bf q},E) \propto \frac{2E_{\bf q}^2\Gamma E}{(E_{\bf q}^2 - E^2)^2 + 4\Gamma^2E^2},\label{eq:3}
\end{equation}
in which $E_{\bf q} = \hbar[(v_{\parallel}q_{\parallel})^2+(v_{\perp}q_{\perp})^2]^{1/2}$ is an anisotropic dispersion with velocities $v_{\parallel}$ and $v_{\perp}$ in the longitudinal and transverse directions relative to ${\bf q}_{\rm m} = (\pi,0)$, $\Gamma = \gamma E$ is the inverse lifetime, and $\bf q$ is the spin-wave wave vector.  $\chi''({\bf q},E)$ does not vary with $q_z$, and is repeated in 2D momentum space with the periodicity of the 2D magnetic wave vector ${\bf q}_{\rm m}$. We fitted the model to the constant-energy cuts allowing $\gamma$, $v_{\parallel}$, $v_{\perp}$, an intensity scale factor and a flat background to vary.
The experimental $Q$ resolution was included.

The parameters determined from the fit are $\gamma = 0.13 \pm 0.06$, $v_{\parallel}=460 \pm 120$\,meV{\AA} and $v_{\perp}=150 \pm 20$\,meV{\AA}. The anisotropic velocity obtained from this analysis is statistically significant. Spectra simulated with the best-fit parameters are shown in Fig.~\ref{fig3}(b) \cite{footnote1}. The simulations match the peak at $Q = 1.2$\,{\AA}$^{-1}$ and closely reproduce the observed dispersion of the signals centered near 2.6, 3.5 and 4.5\,{\AA}$^{-1}$. The model parameters are similar to those found for Ba(Fe$_{1-x}$Co$_x$)$_2$As$_2$: $\gamma = 0.15$, $v_{\parallel}=580$\,meV{\AA}, $v_{\perp}=230$\,meV{\AA} \cite{LES10}. We also considered a model for purely diffusive spin dynamics \cite{TUC12}. The diffusive model also fits the data satisfactorily and leads to the same conclusions as the damped spin wave model \cite{footnote1}.

Despite the limitations inherent in powder-averaging, the success of the model in accounting for features in the data over several Brillouin zones places a tight constraint on the wave vector ${\bf q}_{\rm m}$ that describes the dominant mode of paramagnon excitations in FeSe. As a test of this, we carried out fits with the damped spin-wave model modified to have ${\bf q}_{\rm m} = (\pi,\xi)$ and an isotropic dispersion. The fits with this model were in all cases inferior to those with ${\bf q}_{\rm m} = (\pi,0)$ and an anisotropic dispersion. The best agreement was achieved with $\xi \approx \pi/4$ and is shown in Fig.~\ref{fig3}(c), but the leading edge of the fitted signal near $2.6$\,{\AA}$^{-1}$ is at too low $Q$ compared with the $(\pi,0)$ model reflecting the difference between the magnitude of the wave vector $(\pi,2\pi)$, $Q=2.64$\,{\AA}$^{-1}$, and $(\pi,7\pi/4)$, $Q=2.38$\,{\AA}$^{-1}$ --- see Fig.~\ref{fig3}(a) and \cite{footnote1}.

In this experiment we were unable to cool the sample below 8\,K, and so did not study the magnetic signal in the superconducting state at low energies where a spin resonance could be expected. Instead, we investigated the influence of the structural transition on the magnetic response by performing runs with $E_{\rm i} = 50$\,meV at temperatures of 8, 17, 67 and 104\,K. Figure~\ref{fig4} shows $Q$ cuts through the $(\pi,0)$ position at each temperature. The data are averaged over the energy interval from 11 to 14\,meV to stay within the window where phonon scattering is weak. The magnetic peaks show very little variation with temperature. To quantify this, we fitted a Gaussian function on a linear background to each cut. To within the fitting error the integrated intensity remains constant at $0.10 \pm 0.01\,\mathrm{mb}\,\mathrm{sr}^{-1}$\,meV$^{-1}$\,{\AA}$^{-1}$\,f.u.$^{-1}$, which compares with the value $0.08 \pm 0.01\,\mathrm{mb}\,\mathrm{sr}^{-1}$\,meV$^{-1}$\,{\AA}$^{-1}$\,f.u.$^{-1}$ found at the same energy for LiFeAs at $T=20$\,K $>T_{\rm c}$ \cite{TAY11}. This shows that the spin fluctuations in FeSe have a similar strength to those in other Fe-based superconductors.

The fact that the magnetic response shows very little or no change on crossing the structural phase transition implies that the structural transition is not driven by magnetic fluctuations at the frequencies probed in our experiment. Further, the lack of any change over the entire temperature range studied implies that the paramagnetic moment is constant below 104\,K, in contrast with the notion of a gradual spin-state transition proposed to explain thermally-induced phonon anomalies observed in Raman spectra \cite{GNE13}.

 This study establishes that the collective spin fluctuations in FeSe share many similarities with those in the high-$T_{\rm c}$ Fe arsenide superconductors, including a very steep dispersion and a low frequency response that is strongest at or very close to the square lattice wave vector $(\pi,0)$. We find no direct evidence for competing magnetic orders, although the highly anisotropic spin-wave velocity implies a greater tendency for transverse spin fluctuations. If spin fluctuations are important for the pairing mechanism in Fe-based superconductors then our results show that the ingredients for high-$T_{\rm c}$ are present in FeSe, and something other than conventional magnetic dipole fluctuations must compete with superconductivity. Several different nematic degrees of freedom that could suppress superconductivity have been discussed recently \cite{BAE14,NAK14,SHI14,BOE15,CAO14,WAN15,GLA15,YU15}, and  experiments to search for possible orbital and spin nematic order parameters compatible with $(\pi,0)$ spin fluctuations will be an important next step.

{\it Note added}: Recently, an eprint appeared reporting neutron scattering measurements of the low-energy response in FeSe \cite{WANG15}.

 We thank A. Daoud-Aladine for help with the neutron powder diffraction measurements at ISIS, and M. D. Watson and R. M. Fernandes for discussions. The work was supported by the UK Engineering \& Physical Sciences Research Council (grant EP/I017844). SJS acknowledges the support of a DFG Fellowship (SE2324/1-1).

\bibliography{FeSeBib}

\cleardoublepage

\onecolumngrid

\appendix

\begin{center}

\Large
Supplemental Material:

\vspace{1cm}

\large
{\bf Strong $(\pi,0)$ spin fluctuations in $\beta$-FeSe observed by neutron spectroscopy}

\vspace{1cm}

\normalsize

M. C. Rahn,$^1$ R. A. Ewings,$^2$ S. J. Sedlmaier,$^3$ S. J. Clarke,$^3$ and A. T. Boothroyd$^1$

\vspace{0.2cm}
\small

$^1${\it Department of Physics, University of Oxford, Clarendon Laboratory, Oxford, OX1 3PU, United Kingdom}\\[1pt]
$^2${\it ISIS Facility, STFC Rutherford Appleton Laboratory, Harwell Oxford, Didcot, OX11 0QX, United Kingdom}\\[1pt]
$^3${\it Department of Chemistry, University of Oxford, Inorganic Chemistry Laboratory, Oxford, OX1 3QR, United Kingdom}\\

\end{center}
\vspace{1cm}

\twocolumngrid

\section{Damped harmonic oscillator model}

The damped harmonic oscillator (DHO) model is defined in Eqs.~(\ref{eq:1})--(\ref{eq:3}) of the main article. It is a phenomenological model for damped spin waves. The model has four variable parameters: the longitudinal and transverse spin-wave velocities $v_{\parallel}$ and $v_{\perp}$, the damping parameter $\gamma$, and an intensity scale factor. This model fits the experimental spectrum well, as demonstrated via the $Q$ cuts shown in Fig.~\ref{fig3}(b) of the main article. The fit shows that the spin-wave velocity is anisotropic with $v_{\parallel} \approx 3v_{\perp}$. Figure~S1 (left panels) shows intensity maps for the best-fit DHO model simulated over the same range of energy and wave vector covered by the measurements in Fig.~\ref{fig2} of the main article.

We also consider the possibility that the spin fluctuations are associated with a characteristic wave vector ${\bf q}_{\rm m} = (\pi,\xi)$,  as indicated by some theoretical models.  To perform a simple test we modified the DHO model to have ${\bf q}_{\rm m} = (\pi,\xi)$ and fitted the model to the data for $0 \le \xi \le \pi/2$, constraining for simplicity the spin-wave velocity to be isotropic.  Figure~\ref{fig3}(c) in the main article shows the best fit that could be achieved with this modified DHO model, which was with $\xi \approx \pi/4$. The fit is not as good as the DHO model with anisotropic velocity and ${\bf q}_{\rm m} = (\pi,0)$. The discrepancies are most noticeable for the signal near $2.6$\,{\AA}$^{-1}$ at low energies. The reason for this is illustrated in Fig.~\ref{fig3}(a) of the main article. The unmodified DHO model converges at low energies to ${\bf q}_{\rm m} = (\pi,0)$ and equivalent points in 2D momentum space, whereas the modified DHO model dispersion converges to two points at ${\bf q}_{\rm m} = (\pi,\pm \pi/4)$, and equivalent points. Therefore, in a powder-averaged spectrum at low energies the anisotropic model will show a rise in intensity at $Q=2.64$\,{\AA}$^{-1}$ corresponding to ${\bf q}_{\rm m} = (\pi,2\pi)$, whereas the isotropic model will show a rise in intensity at $Q=2.38$\,{\AA}$^{-1}$ corresponding to ${\bf q}_{\rm m} = (\pi,7\pi/4)$. To illustrate this effect more directly we show in Fig.~S2 the distribution in intensity in 2D momentum space for the two models calculated for the same energies as the $Q$ cuts in Fig.~\ref{fig3}.
\begin{figure}
\includegraphics[width=1.0\columnwidth,trim= 0pt 0pt 0pt 0pt, clip]{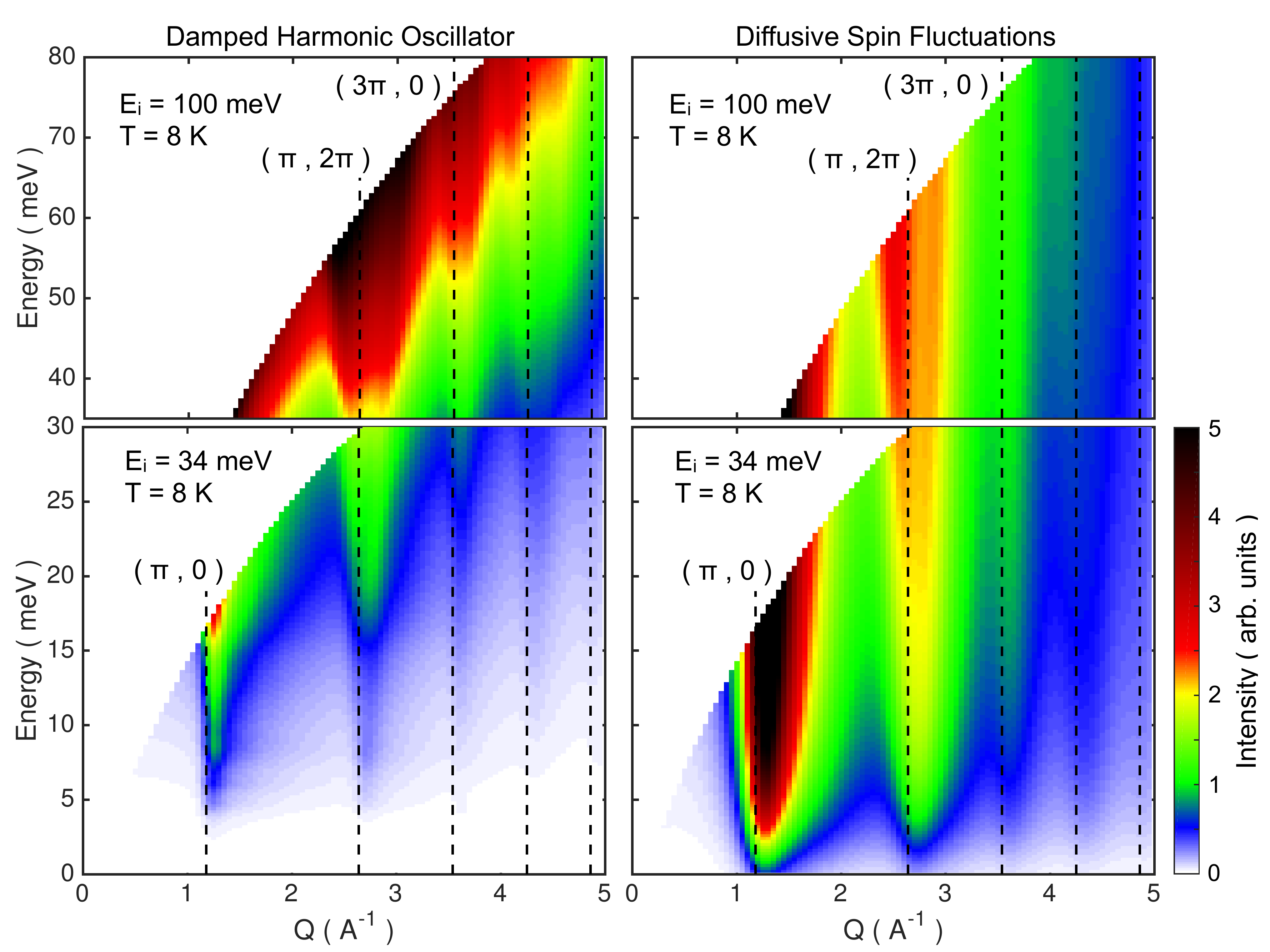}
\\[5pt]
\justify{\label{fig:S1} Fig.~S1. Intensity maps of the best-fit DHO model (left panels) and diffusive model (right panels). The spectra cover the same range of energy and wave vectors as in Fig.~\ref{fig2} of the main article.}
\end{figure}

\begin{figure}
\includegraphics[width=1.0\columnwidth,trim= 0pt 0pt 0pt 0pt, clip]{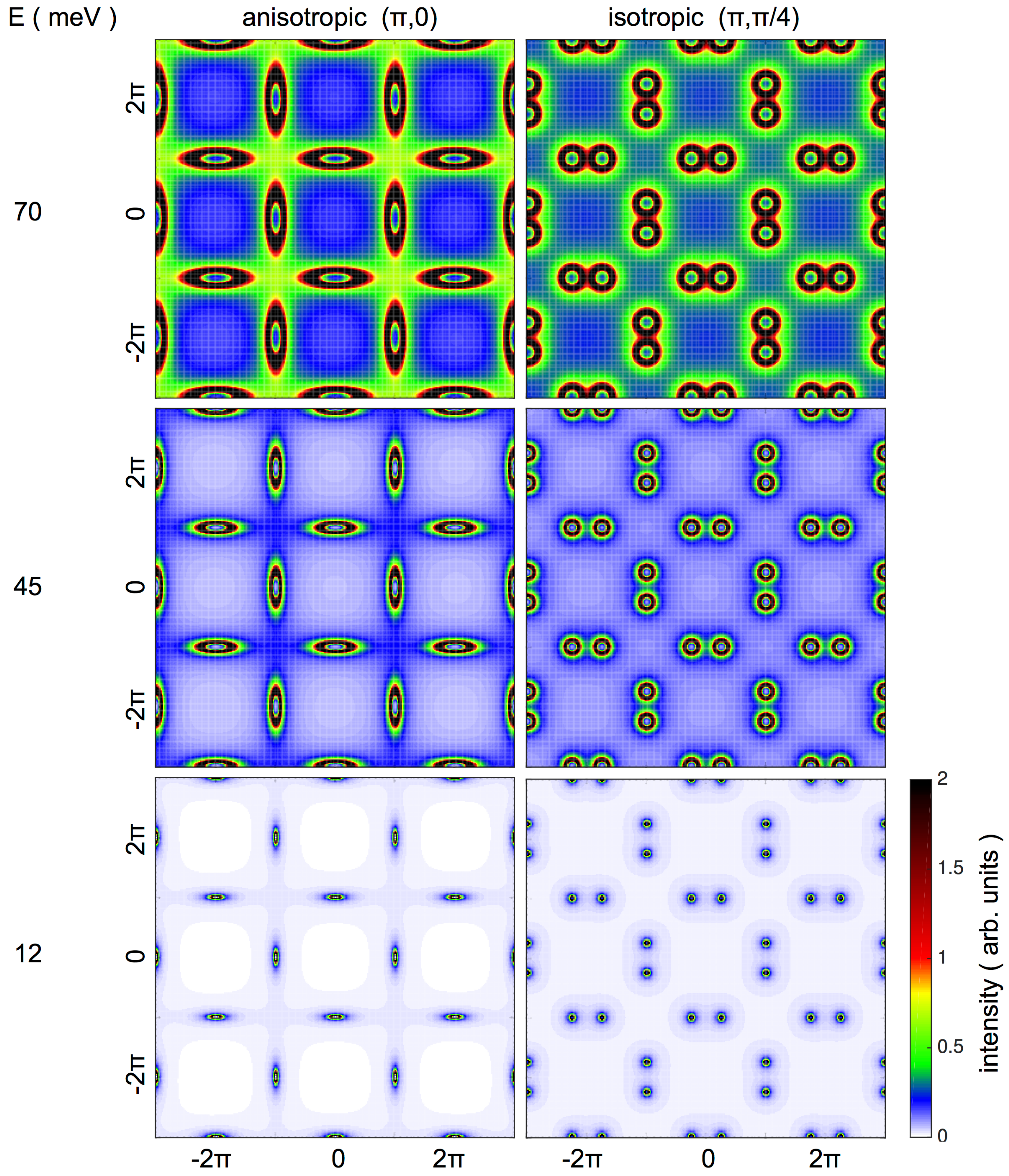}
\\[5pt]
\justify{\label{fig:S1} Fig.~S2. Intensity maps of the DHO model in 2D momentum space at different energies. The magnetic wave vector is ${\bf q}_{\rm m} = (\pi,0)$ in the left panels (best-fit model), and ${\bf q}_{\rm m} = (\pi,\pm \pi/4)$ in the right panels.}
\end{figure}

\vspace{0.6cm}

\section{Diffusive model for correlated paramagnet}

Strictly speaking, the DHO spin-wave model applies to systems with short-range magnetic order. The justification for applying it to FeSe is that there are strong magnetic correlations, which means that the spins will appear ordered over sufficiently short distances and timescales. Therefore, above some crossover energy scale the spectrum is expected to be very similar to that of a system whose spins are ordered with propagation vector ${\bf q}_{\rm m}$ and whose excitations are propagating damped spin waves.

An alternative approach is to employ a description that does not imply any magnetic order in the ground state. Such a phenomenological model was used by Tucker {\it et al.} to analyse neutron scattering data on superconducting Ba(Fe$_{1-x}$Co$_x$)$_2$As$_2$  with $x = 0.047$ \cite{TUC12}. This model is designed to capture the diffusive nature of the spin dynamics inherent in a paramagnet while at the same time allowing spatial and temporal magnetic correlations. The scattering intensity is proportional to the imaginary part of the generalized susceptibility, see Eqs.~(\ref{eq:1})--(\ref{eq:2}), which in the diffusive model is given by
\begin{equation}
\chi''({\bf q},E) \propto \frac{\Gamma_{\bf q}E}{E^2 + \Gamma_{\bf q}^2(1+\xi_{\bf q}^2 q^2)^2},\label{eq:4}
\end{equation}
where the momentum-dependent correlation lengths and relaxation rates are
\begin{eqnarray}
\xi_{\bf q}^2 q^2 & = & \xi_{\parallel}^2q_{\parallel}^2 + \xi_{\perp}^2q_{\perp}^2\label{eq:5}\\[5pt]
\Gamma_{\bf q} & = & \Gamma + \frac{\alpha^2}{\Gamma}\left(\Gamma_{\parallel}^2q_{\parallel}^2+\Gamma_{\perp}^2q_{\perp}^2\right).\label{eq:6}
\end{eqnarray}
The model has six variable parameters including an intensity scale factor. The parameters used here are consistent with those defined by Tucker {\it et al.} \cite{TUC12}. Specifically, the parameters $\xi$, $\Gamma$, $\eta_{\xi}$ and $\eta_{\Gamma}$ in Tucker {\it et al.} are given by
\begin{eqnarray}
\xi_{\parallel}^2 & = & \xi^2(1+\eta_{\xi}),\hspace{15pt} \xi_{\perp}^2  =  \xi^2(1-\eta_{\xi}),\\[5pt]
\Gamma_{\parallel}^2 & = & \Gamma^2(1+\eta_{\Gamma}),\hspace{15pt} \Gamma_{\perp}^2  =  \Gamma^2(1-\eta_{\Gamma}).
\end{eqnarray}
In Eqs.~(\ref{eq:4})--(\ref{eq:6}), ${\bf q} = (q_{\parallel}, q_{\perp})$ is the reduced wave vector measured relative to a magnetic wave vector, i.e.~${\bf q}=0$ is at ${\bf q}_{\rm m}$. The components $q_{\parallel}$ and $q_{\perp}$ are parallel and perpendicular to the vector joining ${\bf q}_{\rm m}$ to the nearest reciprocal lattice vector, e.g. for ${\bf q}_{\rm m}=(\pi,0)$, $q_{\parallel}$ and $q_{\perp}$ are components parallel and perpendicular to ${\bf q}_{\rm m}$. At low energies this model describes an elliptical intensity distribution centred on ${\bf q}=0$, with $q_{\parallel}$ and $q_{\perp}$ as the principal axes of the ellipse. $\chi''({\bf q},E)$ is repeated in 2D momentum space at each of the 2D wave vectors ${\bf q}_{\rm m}$.

The best fit of the diffusive model to the data was obtained with parameters $\xi_{\parallel} = 7 \pm 2$\,\AA, $\xi_{\perp} = 1.5 \pm 0.8$\,\AA, $\Gamma_{\parallel} = 0$\,meV, $\Gamma_{\perp} = 7 \pm 5$\,meV and $\alpha = 7 \pm 5$\,\AA. The parameter $\Gamma_{\parallel}$ was not well controlled and had little effect on the fit. The large difference between $\xi_{\parallel}$ and $\xi_{\perp}$ means that the intensity distribution is highly anisotropic, consistent with what was found with the DHO model.

Spectra simulated with the best-fit parameters of the diffusive model are shown in Fig.~S3. The simulated $Q$ cuts match the data well when ${\bf q}_{\rm m}=(\pi,0)$ (left panel), but less well with ${\bf q}_{\rm m} = (\pi,\pm \pi/4)$ and an isotropic intensity distribution (right panel).

The main conclusions are, firstly, that the phenomenological diffusive model provides a good description of the magnetic dynamics of FeSe, and second, that the analysis with the diffusive model reinforces the findings from the DHO spin-wave model analysis that FeSe is close to an instability towards $(\pi,0)$ antiferromagnetism and has highly anisotropic magnetic correlations.

\begin{figure}[b]
\includegraphics[width=1.0\columnwidth,trim= 0pt 0pt 0pt 0pt, clip]{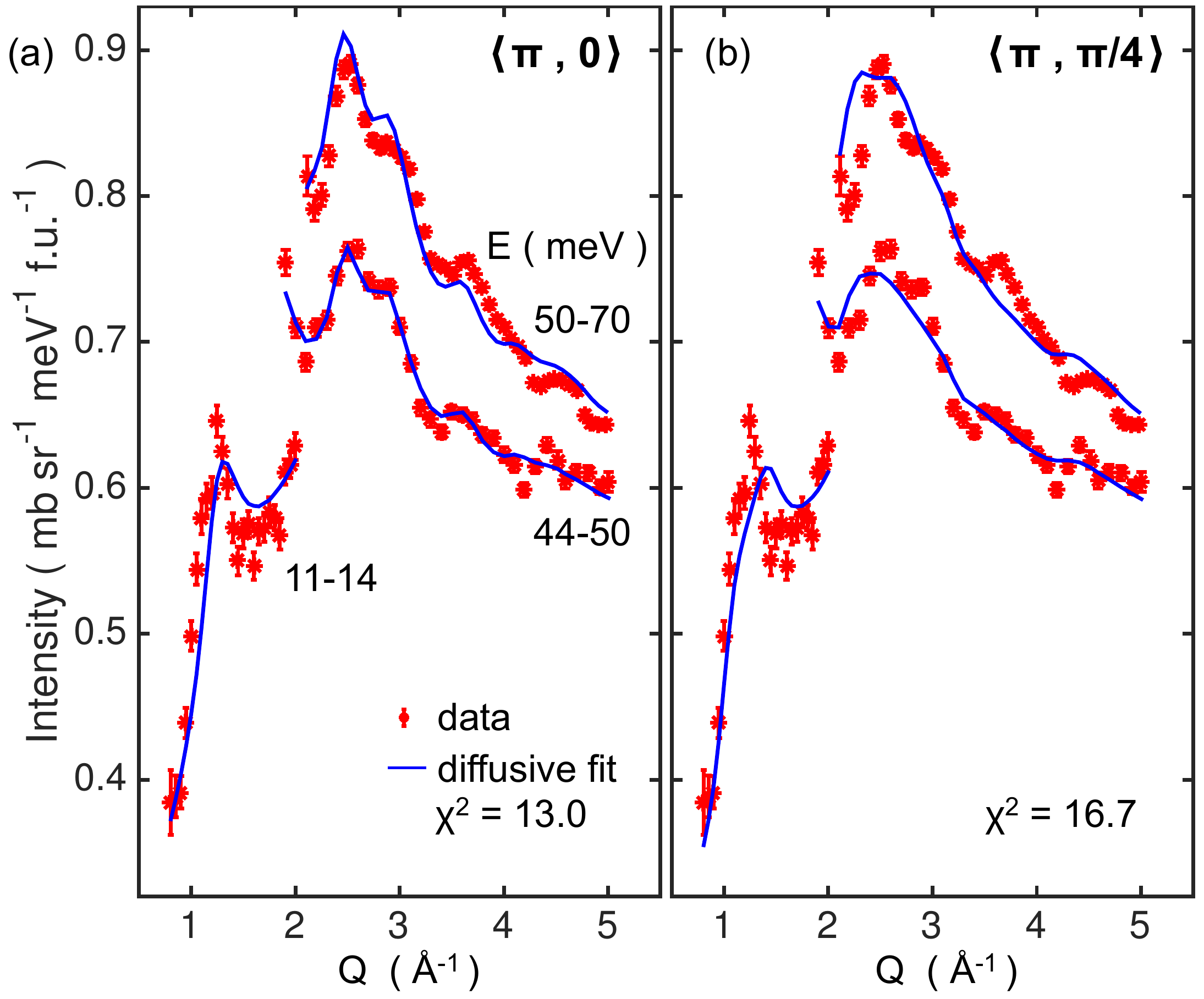}
\\[5pt]
\justify{\label{fig:S3} Fig.~S3. Constant-energy cuts through the data together with fits with the diffusive model. (a) ${\bf q}_{\rm m}= (\pi,0)$, (b) ${\bf q}_{\rm m} = (\pi,\pi/4)$ and an isotropic dispersion.}
\end{figure}

\end{document}